\documentclass[prl,twocolumn]{revtex4-1}
\usepackage[utf8]{inputenc}
\usepackage{amsmath}
\usepackage{amsfonts}
\usepackage{amssymb}
\usepackage{bm}
\usepackage{graphicx}
\usepackage{xcolor}
\usepackage{microtype}

\begin{document}
\title{Interacting Hofstadter Interface}
\author{Bernhard Irsigler}
\author{Jun-Hui Zheng}
\author{Walter Hofstetter}
\affiliation{Institut f{\"u}r Theoretische Physik, Goethe-Universit{\"a}t, 60438
Frankfurt/Main, Germany}
\begin{abstract}
Two-dimensional topological insulators possess conducting edge states at their boundary while being insulating in the bulk. We investigate the edge state emergent at a smooth topological phase boundary of interacting fermions within a full real-space analysis of the time-reversal invariant Hofstadter-Hubbard model. We characterize the localization of the edge state and the topological phase boundary by means of the local compressibility, the spectral density, a generalized local spin Chern marker as well as the Hall response and find good agreement between all these quantities. Computing the edge state spectra at the interface we observe robustness of the edge state against fermionic two-body interactions and conclude that interactions only shift its position. Hence the bulk-boundary correspondence for the interacting system is confirmed. Since experimental probing of edge states remains a challenge in ultracold atom setups, we propose the detection of the local compressibility by measuring correlations with a quantum gas microscope.
\end{abstract}
\maketitle

Since the realizations of topologically non-trivial states with cold atoms including the bosonic Hofstadter model \cite{Aidelsburger2013,Miyake2013}, the fermionic Haldane model \cite{Jotzu2014}, and the measurements of the Chern number \cite{Aidelsburger2014} and the Berry curvature \cite{Flaschner2016}, much attention has been paid to drawing parallels between topological insulators in condensed matter and cold atoms systems. One prominent example is the detection of edge states, which has been succesfully performed with scanning tunneling microscopy at a the edge of a crystalline single-layer \cite{Wu2016,Li2016}, as well as in photonic graphene \cite{Rechtsman2013} and even in classical systems \cite{Susstrunk2015}. From the cold atom perspective topological edge states have been observed in synthetic dimensions \cite{Mancini2015,Stuhl2015} and chiral Meissner-like currents in ladder systems \cite{Atala2014}. The smooth confinement of cold atom setups, however, makes it difficult to observe edge states in full two-dimensional systems since the inherent inhomogeneity blurs the differentiation of edge and bulk \cite{Stanescu2010,Buchhold2012,Galilo2017}. Proposals for detecting edge states in cold atom setups feature Bragg scattering \cite{Goldman2012,Buchhold2012} and wavepacket dynamics \cite{Goldman2013}. Another approach is to shift the topological phase boundary, which is usually at the system's boundary, to the center of the system by means of topological interfaces, which have been studied theoretically \cite{Reichl2014,Goldman2016} and experimentally in one dimension \cite{Leder2016,Meier2016,An2017}.

Here, we extend the idea of topological interfaces to a spinful, two-dimensional, fermionic system and are particularly interested in the influence of finite two-body interactions. To this end we make use of an inhomogeneous superlattice. We characterize the topological interface locally by computing the closing of the gap via the compressibility and the spectral density, the edge states, the Hall pumping behavior, and the local spin Chern marker. We hence confirm the bulk-boundary correspondence for interacting systems. The influence of interactions on known topological phases such as the Quantum Spin Hall state is of fundamental interest, both for topological electronic materials \cite{Xiao2010} and for quantum simulators based on ultracold gases in synthetic gauge fields \cite{Hofstetter2018,Rachel2018}.

\begin{figure}[ht]
\includegraphics[width=\columnwidth]{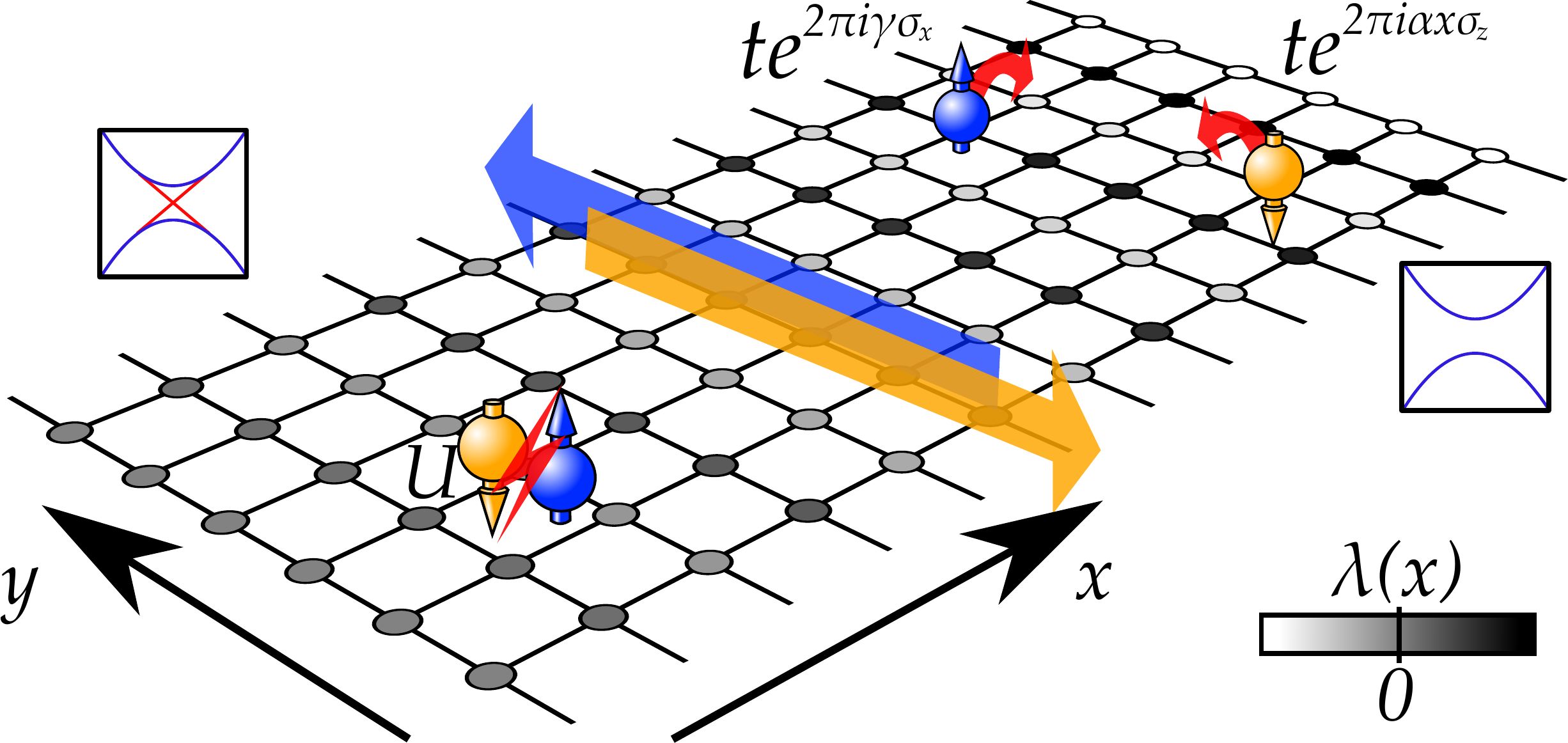}
\caption{Schematic setup: Hamiltonian \eqref{hamiltonian} in cylinder geometry. See text for details.}
\label{schem}
\end{figure}

The time-reversal invariant version of the Hofstadter Hamiltonian \cite{Goldman2010} has been investigated in the presence of two-body interactions for infinite systems \cite{Cocks2012,Orth2013,Kumar2016}:
\begin{equation}
\hat{H} = \hat{H}_0 + \hat{H}_\lambda + \hat{H}_\text{int},
\label{hamiltonian}
\end{equation}
where
\begin{equation}
\hat{H}_0=
-t\sum_{\bm{j}}\left[\bm{\hat{c}}_{\bm{j}+\hat{x}}^\dag e^{i\theta_x} \bm{\hat{c}}_{\bm{j}}+\bm{\hat{c}}_{\bm{j}+\hat{y}}^\dag e^{i\theta_y} \bm{\hat{c}}_{\bm{j}}\right]+\rm{H.c.}
\label{htri}
\end{equation}
represents non-interacting fermions in a two-dimensional lattice exposed to the gauge fields $\theta_x=2\pi \gamma\sigma_x$ and $\theta_y=2\pi \alpha x\sigma_z$, with $\sigma_i$ being the $i$-th Pauli matrix acting in spin space. In Eq.~\eqref{htri} $\bm{\hat{c}}_{\bm{j}}=\left(\hat{c}_{\bm{j},\uparrow},\hat{c}_{\bm{j},\downarrow}\right)$ denotes the annihilation operator in spin space for lattice site $\bm{j}=(x,y)$, $t$ is the hopping amplitude which we set to 1, $\gamma$ is the spin-mixing amplitude of the spin-orbit coupling, and $\alpha$ is the plaquette flux which we set to 1/6. The second term 
\begin{equation}
\hat{H}_\lambda =\sum_{\bm{j}} \lambda(x) \bm{\hat{c}}_{\bm{j}}^\dag\bm{\hat{c}}_{\bm{j}}
\label{hlambda}
\end{equation}
is an additional on-site staggering potential with amplitude $\lambda(x)=\lambda_x(-1)^x$ in its original form \cite{Goldman2010} with $\lambda_x$ being a constant.  For the non-interacting case there exists a topological phase below the critical staggering potential $\lambda_c=1.25$. This value increases with increasing interaction strength $U$ \cite{Kumar2016}. In order to create a topological phase boundary within the system we linearly increase the staggering potential as a function of position
\begin{equation}
\lambda(x)=\left[\lambda_L+(\lambda_R-\lambda_L)\frac{x}{M}\right](-1)^x,
\label{staggering}
\end{equation}
such that it has the value $\lambda_c$ in the center of the system for the non-interacting case.  The system we investigate is schematically depicted in Fig.~\ref{schem}. The staggering parameters $\lambda_L$ and $\lambda_R$ in Eq.~\eqref{staggering} denote the amplitude of the staggering potential at the left edge $x=0$ and the right edge $x=M$ of the system, respectively. With optical lattices the potential \eqref{staggering} would in principle be realizable by using beat amplitudes with two interfering, slightly detuned laser beams. The last term in Eq.~\eqref{hamiltonian}
\begin{equation}
\hat{H}_\text{int}=U\sum_{\bm{j}}\hat{n}_{\bm{j},\uparrow}\hat{n}_{\bm{j},\downarrow}
\label{hint}
\end{equation}
is the fermionic interaction term, where $\hat{n}_{\bm{j},\sigma}=\hat{c}_{\bm{j},\sigma}^\dag \hat{c}_{\bm{j},\sigma}$ is the local density operator of the spin-$\sigma$ fermions and $U$ is the on-site interaction strength. 

Interaction effects are most pronounced at half-filling and we are mainly interested in insulating phases, both topologically trivial and non-trivial. Therefore, we need the system to be gapped at half-filling. For the non-interacting case this can be achieved  either by a finite next-nearest neighbor hopping term leading to the Harper-Hofstadter-Hatsugai Hamiltonian \cite{Hatsugai1990} or via finite spin-orbit coupling $\gamma>0$ and staggering potential $\lambda>0$ \cite{Orth2013}. Here we follow the latter scenario for maximal spin-orbit coupling $\gamma=1/4$, since this yields the largest gap at half-filling. It allows us to transform the spin degrees of freedom from strongly coupled physical spins to decoupled virtual spins. We apply the transformation
\begin{equation}
{\hat{d}}_{\bm{j},\sigma}=
\begin{cases}
{\hat{c}}_{\bm{j},\sigma},\text{ if }x\text{ is even}\\
{\hat{c}}_{\bm{j},\bar{\sigma}},\text{ if }x\text{ is odd},
\end{cases}
\label{trafo}
\end{equation}
with $\bar{\uparrow}=\downarrow$ and $\bar{\downarrow}=\uparrow$. The Hamiltonian \eqref{htri} then acquires the form
\begin{equation}
\hat{H}_0=
-t\sum_{\bm{j}}\left[\bm{\hat{d}}_{\bm{j}+\hat{x}}^\dag e^{i\theta_x}\bm{\hat{d}}_{\bm{j}}+\bm{\hat{d}}_{\bm{j}+\hat{y}}^\dag e^{i\theta_y} \bm{\hat{d}}_{\bm{j}}\right]
+\mathrm{H.c.}
\end{equation}
with $\theta_x=0$ and $\theta_y=2\pi(-1)^x\alpha x\sigma_z$ being spin diagonal. Note that under this transformation the spin mixing vanishes completely and the flux becomes staggered $\alpha\rightarrow(-1)^x\alpha$, which breaks the spatial homogeneity of the plaquette flux pattern. Since they are local the terms \eqref{hlambda} and \eqref{hint} are invariant under this transformation, i.e. $\bm{\hat{c}}_{\bm{j}}$ is substituted by $\bm{\hat{d}}_{\bm{j}}$. The decoupled virtual spins carry the same topological properties as the physical spin system as we show in the supplemental material \cite{supp}.

Since our systems of interest are highly inhomogeneous we make use of the real-space version of Dynamical Mean Field Theory (RDFMT) \cite{Snoek2008}. In RDMFT the full lattice many-body problem is reduced to solving a single impurity problem for each lattice site, with these impurity problems being coupled via the lattice Dyson equation. The only approximation therein is made by assuming the selfenergy to be diagonal in real space $\Sigma^{\sigma\sigma'}_{\bm{ij}}=\Sigma^{\sigma\sigma'}_{\bm{ii}}\delta_{\bm{ij}}$. To solve the impurity problem we use a continuous-time Monte-Carlo solver \cite{Gull2011}. Our lattices consist of 48$\times$48 sites with open boundary conditions (OBC) in the $x$-direction and periodic boundary conditions (PBC) in the $y$-direction referring to the cylinder geometry. Therefore $k_y$ is a good quantum number. Experimentally, cylinder geometries have recently been realized in cold atom setups \cite{Han2018,Li2018}.

\begin{figure}[ht]
\includegraphics[width=\columnwidth]{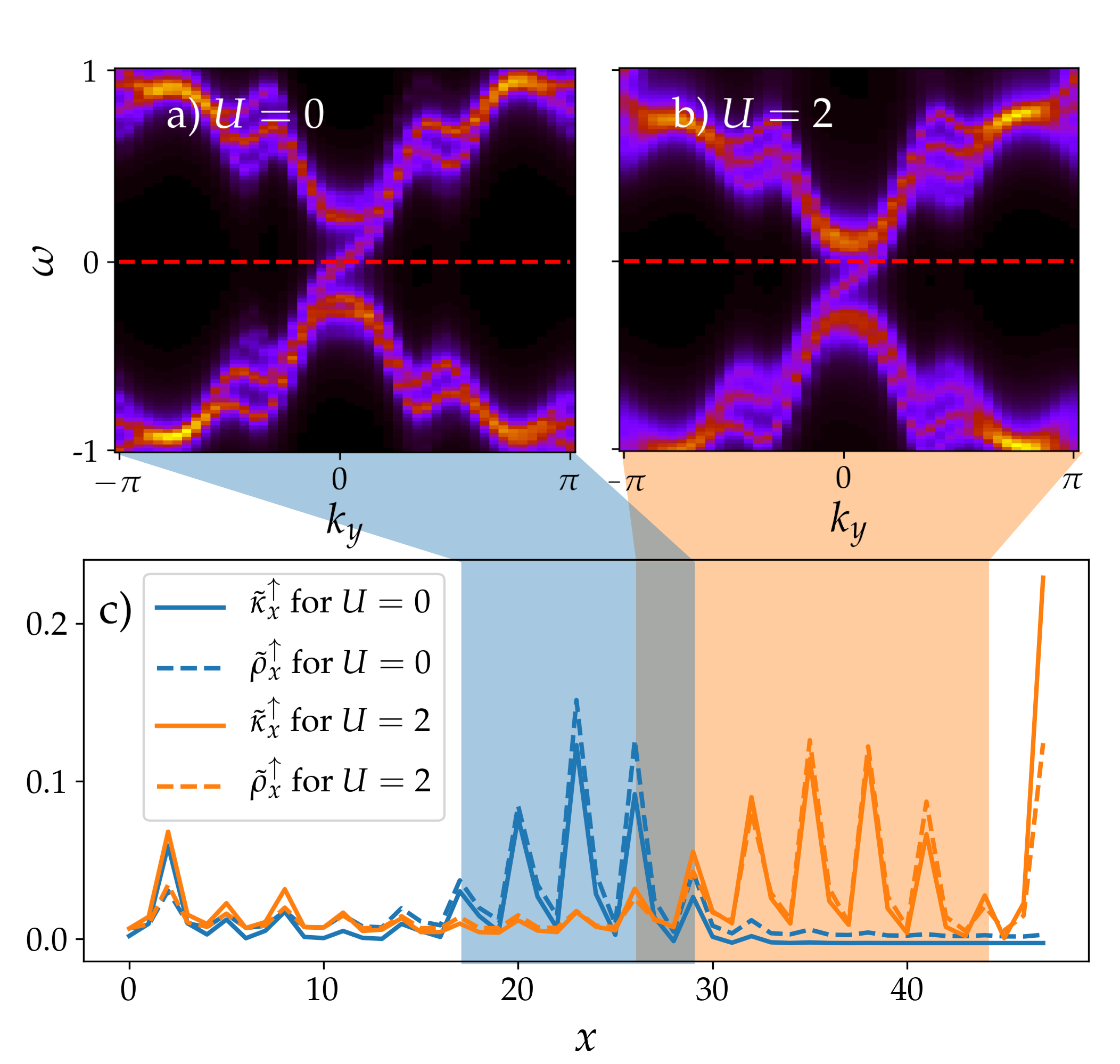}
\caption{Subfigure c): momentum-integrated spectral density $\tilde{\rho}_x^\sigma$ and local compressibility $\tilde{\kappa}_x^\sigma$ in the virtual spin basis \eqref{trafo} as functions of the lattice site index $x$ for interaction strengths $U=0$ and $U=2$ and staggering parameters $\lambda_L=0$ and $\lambda_R=2.5$. Edge state of the interface shown in a) for $U=0$, computed from the non-interacting spectral function of the blue region marked in c), and in b) for $U=2$, computed from the analytically continued interacting spectral function of the orange region in c).}
\label{spco}
\end{figure}

For the detection of the gapless interface a $x$-position-resolved observable is required. One choice is the $y$-momentum-integrated spectral density at the Fermi level $\rho_x^\sigma\equiv\int\mathrm{d}k_y\rho_x^\sigma(\omega=0,k_y)$, where the spectral density is
\begin{equation}
\rho^\sigma_x(\omega,k_y)=-\frac{1}{\pi}\mathrm{Im}G_{xx}^{\sigma\sigma}(\omega,k_y),
\label{spec}
\end{equation}
and the single-particle Green's function is
\begin{equation}
G^{\sigma\sigma'}_{xx'}(\omega,k_y)=\left[\frac{1}{\omega+i\delta+\mu-H(k_y)-\Sigma(\omega)}\right]^{\sigma\sigma'}_{xx'}
\label{regr}
\end{equation}
with $\delta$ being a small real number. A further quantity we focus on in this work is the local compressibility \cite{Zhou2011,Drewes2016,Nozadze2016}. A lattice version of the local compressibility reads

\begin{equation}
\kappa_x^\sigma(k_y)=\left.\frac{\partial n^\sigma_{\bm{i}}}{\partial\mu}\right|_{\beta,V}
=\beta\sum_{\bm{j},\sigma'}\left[\langle \hat{n}^\sigma_{\bm{i}} \hat{n}^{\sigma'}_{\bm{j}}\rangle-\langle \hat{n}^\sigma_{\bm{i}}\rangle\langle \hat{n}^{\sigma'}_{\bm{j}}\rangle\right]
\label{comp1}
\end{equation}
where we used the hybrid notation $\bm{i}=(x,k_y)$. Furthermore, $\beta$ is the inverse temperature, $n_{\bm{i}}^\sigma\equiv \langle \hat{n}^\sigma_{\bm{i}}\rangle$, and $\langle\cdot\rangle$ denotes the grand-canonical thermal average. The right-hand-side of Eq.~\eqref{comp1} corresponds to non-local density-density correlations, which are not directly accessible in DMFT calculations. However, an expression of the local compressibility in terms of Matsubara Green's functions $G^{\sigma\sigma'}_{\bm{ij}}(i\omega_n)$, with $\omega_n=\pi(2n+1)/\beta$, is derived in the supplemental material \cite{supp}:
\begin{equation}
\kappa_x^\sigma(k_y)\approx-\frac{1}{\beta}\sum_{n,x',\sigma'}G^{\sigma\sigma'}_{xx'}(i\omega_n,k_y)G^{\sigma'\sigma}_{x'x}(i\omega_n,k_y),
\label{comp3}
\end{equation}
where we assumed that the selfenergy depends only weakly on the chemical potential $\partial \Sigma/\partial \mu \ll 1$. The right-hand-side of Eq.~\eqref{comp1} should be accessible through the statistical evaluation of quantum gas microscope images \cite{Tai2017}. Therefore the local compressibility could in principle be measured in experiments, revealing information about gapless edge states or interfaces.

In Fig.~\ref{spco}~c) we show the momentum-integrated spectral density $\tilde{\rho}_x^\sigma$ and local compressibility $\tilde{\kappa}_x^\sigma$, in the virtual spin basis \eqref{trafo} denoted by the tilde, as functions of the lattice site index $x$ for $\lambda_L=0$ and $\lambda_R=2.5$ for the non-interacting case $U=0$ as well as for moderate interaction strengths $U=2$. We observe that both quantities agree, in contrast to the superconducting case in Ref.~\cite{Nozadze2016}, and that deviations emerge through the choice of a small but finite broadening parameter $\delta$ in Eq.~\eqref{regr}. This is discussed in the supplemental material \cite{supp}. For $U=0$ we observe a peak of the compressibility in the center of the system where $\lambda(x)=\lambda_c$ as expected, corresponding to the edge state of the interface which we show in Fig.~\ref{spco}~a) by plotting the spectral density $\tilde{\rho}_x^\sigma(\omega,k_y)$ for the region highlighted in blue in c). For $U=2$ the interface is shifted to the right, i.e. to higher values of the staggering amplitude, as we expect from Ref.~\cite{Kumar2016}. Here the edge state in Fig.~\ref{spco}~b) is computed via the Maximum Entropy Method \cite{Jarrell1996,Wang2009} from RDMFT results for the orange region in c). The robustness of the edge state against interactions stems from its topological protection, which can only be lifted if the bulk gap is closed. According to Ref.~\cite{Kumar2016} this is expected when the interactions are large enough such that the system enters a magnetic phase.

\begin{figure}[ht]
\includegraphics[width=\columnwidth]{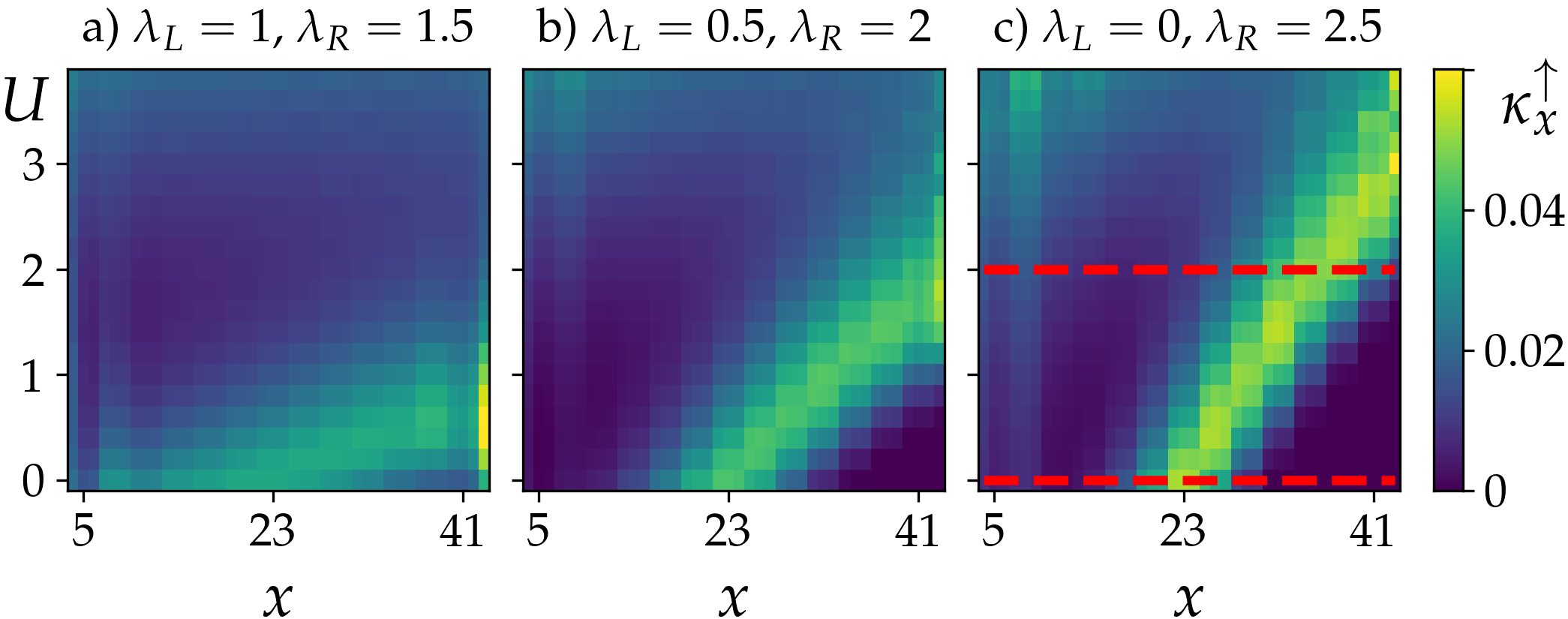}
\caption{Local compressibility computed from RDMFT results as function of the site index $x$ and the interaction strength $U$ for three different sets of staggering parameters $\lambda_L$ and $\lambda_R$. The red dashed lines in c) denote the interaction strengths $U=0$ and $U=2$, which are also used in Fig.~\ref{spco}~c) and Fig.~\ref{pump}.}
\label{comp}
\end{figure}

The spikey profile in Fig.~\ref{spco} c) can be explained through the transformation \eqref{trafo}. Since it maps the hopping amplitude in $y$-direction onto a complex phase proportional to $(-1)^x\alpha$, it does not recover a homogeneous flux pattern as in the original Hofstadter model but rather a position-dependent pattern. The plaquette flux becomes $(2x+1)(-1)^x\alpha$ which, up to a sign, has periodicity 3. The flux per magnetic unit cell, however, is not affected and amounts to 1 as in the original Hofstadter model.

 In Fig.~\ref{comp} we show the local compressibility as function of $U$ for different staggering parameters $\lambda_L$ and $\lambda_R$ from RDMFT results. For better visibility we cut the edges and smoothen the data over three lattice sites. Dark regions correspond to insulators and light regions to gapless states. We observe that the interface becomes more localized with increasing window size $\lambda_R-\lambda_L$. The location of the interface is consistent with the phase diagram in Ref.~\cite{Kumar2016}. Hence, we find a phase separation between the QSH phase and the topologically trivial band insulator in real space separated through a gapless interface.

\begin{figure}[ht]
\includegraphics[width=\columnwidth]{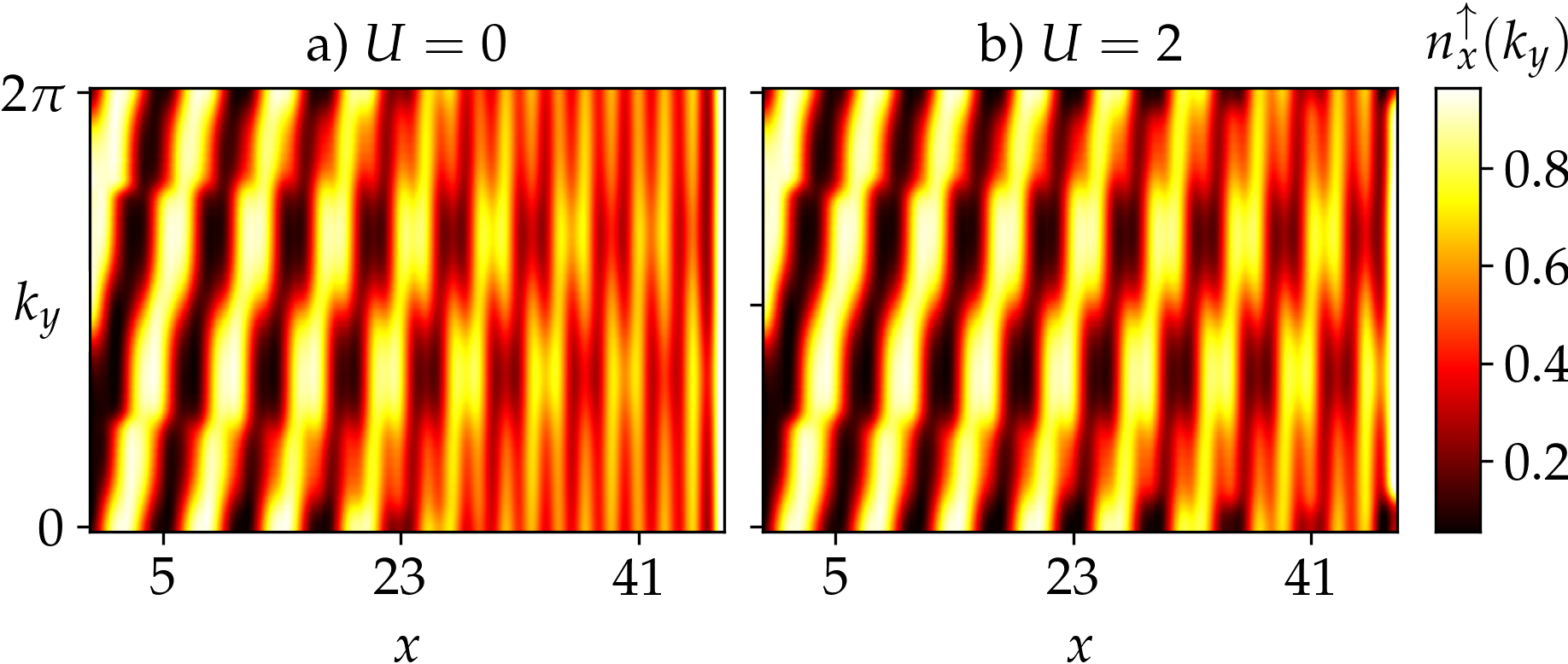}
\caption{Particle density $n_x^\uparrow(k_y)$ as function of site index $x$ and momentum $k_y$ for $\lambda_L=0$, and $\lambda_R=2.5$ for a) $U=0$ as well as b) $U=2$ corresponding to the two red dashed  lines in Fig.~\ref{comp} c).}
\label{pump}
\end{figure}

For a first justification of the topological distinction of the two insulating phases in Fig.~\ref{comp} we determine the Hall response of the system to a bias force $\dot{k}_y$ \cite{Wang2013,Aidelsburger2014,Lohse2015,Nakajima2016}. In Fig.~\ref{pump} we present the $x$-position and $y$-momentum-resolved particle density $n^\sigma_x(k_y)=\int_{-\infty}^0\mathrm{d}\omega\rho_x^\sigma(k_y)$ for $\lambda_L=0$, and $\lambda_R=2.5$ for two interaction strengths $U=0$ and $U=2$ referring to the two dashed red lines in Fig.~\ref{comp} c). A pump cycle corresponds to varying $k_y$ adiabatically by $2\pi$. We observe that after one cycle particles are pumped six lattice sites further in the left half of the system, corresponding to the size of the magnetic unit cell $1/\alpha=6$. In contrast, there is no such pumping in the right half. This confirms a topologically non-trivial state in the left phase and a topologically trivial insulator in the right phase. We show exemplarily the same scenario for the case of interaction energy $U=2$ in Fig.~\ref{pump} b). Here the topologically non-trivial phase is clearly extended to the right, following the trend of Fig.~\ref{comp} c). As stated in Ref.~\cite{Wang2013}, pumping could in principle be measured through a hybrid version of time-of-flight imaging, which is directly applicable to the system discussed here.

Alternatively, we apply the idea of the local Chern marker (LCM) introduced by Ref.~\cite{Bianco2011} for finite systems and heterojunctions. It has been investigated for a spin-decoupled interacting system \cite{Amaricci2017}, as well as for a quantum gas within a quasicrystal \cite{Tran2015}. 
At vanishing spin-orbit coupling $\gamma=0$ a time-reversal invariant system exhibits a vanishing charge Chern number $C_\uparrow+C_\downarrow$ even in topologically non-trivial states. However, the spin Chern number $C_\uparrow-C_\downarrow=2C_\uparrow$ can be finite and indicates the presence of the QSH state. For finite spin-orbit coupling $\gamma>0$ the spin Chern number $C=\sum_{\alpha,\beta}\alpha C_{\alpha,\beta}$ has been proposed \cite{Sheng2006}. However, for large spin-orbit coupling $\gamma\approx1/4$ it fails combined with the LCM method. Moreover the density matrix  method \cite{Zheng2017} fails because the density matrix becomes gapless. In contrast, here we consider the decoupled virtual spin system \eqref{trafo}, and we apply the LCM to a single spin component.

\begin{figure}[ht]
\includegraphics[width=\columnwidth]{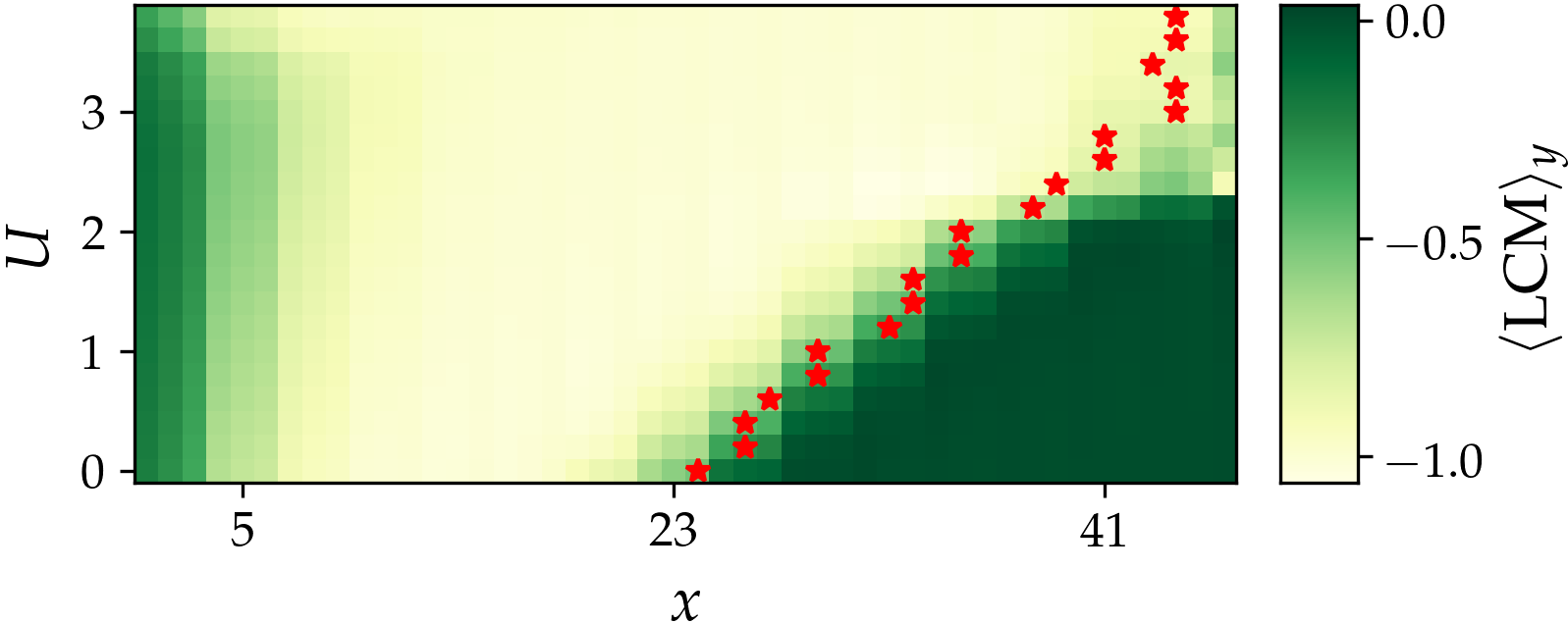}
\caption{Local Chern marker averaged over spatial dimension $y$ and smoothed over three lattice sites along the $x$-direction as function of lattice site $x$ and interaction strength $U$ for $\lambda_L=0$ and $\lambda_R=2.5$. Red stars denote the maxima of the compressibility in Fig.~\ref{comp} c).}
\label{lcm}
\end{figure}

The main point of the LCM is to reidentify the position operator in the definition of the Chern number \cite{Thouless1982} and to omit the trace over the whole system. One is left with a position-dependent marker, which recovers the corresponding Chern number in the bulk value and sums up to zero in finite systems. The marker itself is not quantized. The LCM is computed as follows, with details in the supplemental material \cite{supp}:
\begin{equation}
\mathrm{LCM}=-2\pi i \langle x,y|\left[\hat{P}\hat{x}\hat{P},\hat{P}\hat{y}\hat{P}\right]|x,y\rangle,
\label{LCM}
\end{equation}
where $\hat{P}$ is the projector onto the occupied states and $\hat{x}$ and $\hat{y}$ are the position operators of the respective spatial direction. In order to treat interaction effects on the LCM we make use of the effective topological Hamiltonian approach \cite{Wang2012} $\hat{H}_\text{int}\rightarrow \Sigma(\omega=0)$ and apply the LCM to the effectively non-interacting topological Hamiltonian. In Fig.~\ref{lcm} we present the LCM as function of the lattice site index $x$ and the interaction strength $U$, smoothed over three lattice sites. For $U=0$ we recover a topologically trivial phase $\langle\text{LCM}\rangle_y=0$ on the right-hand-side $x>23$ and a topologically non-trivial phase $\langle\text{LCM}\rangle_y=-1$ on the left-hand-side $5<x<23$. At the left boundary there is a small semi-metallic region \cite{Kumar2016} such that $\langle\text{LCM}\rangle_y$ vanishes. For increasing interaction strengths the topologically non-trivial phase grows as we expect from Fig.~\ref{pump} b). We depict the maxima of the compressibility of Fig.~\ref{comp} c), which correspond to the location of the interface, as red stars in Fig.~\ref{lcm} and find good agreement with the jump of the topological invariant. Finally, we confirm the bulk-boundary correspondence \cite{Hasan2010,Essin2011} for the interacting system by comparing the change in the topological invariant in Fig.~\ref{lcm}, which is equal to 1, and the single edge state which is observed in Figs.~\ref{spco}~a) and b).

In conclusion, we induced a topological phase separation of a time-reversal invariant fermionic system by applying a linearly increasing staggering potential in addition to the lattice potential. The two insulating many-body phases are separated by a gapless interface. We computed the spatially resolved compressibility, which is accessible in experiments and reveals the location of the interface. Moreover, we distinguished the two different insulators by looking at their pumping behavior and by computing a local spin Chern marker. By comparing the topological invariant and the edge state spectrum we confirmed the bulk-boundary correspondence for interacting systems. Finally, we find that interactions in this parameter regime cannot destroy the edge state and only shift its position together with the topological interface.

All necessary experimental techniques required for realizing our proposal exist, however, we are aware of the challenge of combining artificial gauge fields and two-body interactions. Since it is a convenient way of realizing an interacting, time-reversal invariant topological interface with cold atoms, our proposed setup may provide a platform for investigating spin transport properties and non-equilibrium quenches of edge states. Eventually, it should be directly extendable to three dimensions \cite{Scheurer2015}.
\begin{acknowledgements}
We thank Nathan Goldman, Peter Orth, Jaromir Panas, Tao Qin, and Christof Weitenberg for fruitful discussions. This research was funded by the Deutsche Forschungsgemeinschaft (DFG, German Research Foundation) via Research Unit FOR 2414 under project number 277974659. This work was also supported by the Deutsche Forschungsgemeinschaft (DFG) via the high-performance computing center LOEWE-CSC.
\end{acknowledgements}

\bibliographystyle{apsrev4-1}
%

\end{document}


\begin{center}
{\large\bf Supplemental Material for ``Interacting Hofstadter Interface"}

~\\

{Bernhard Irsigler, Jun-Hui Zheng, and Walter Hofstetter}

\textit{Institut f{\"u}r Theoretische Physik, Goethe-Universit{\"a}t, 60438
Frankfurt/Main, Germany}


\end{center}

\section{Number of particles and local compressibility}
We first derive the number of particles in terms of the Matsubara Green's function. Starting with the expression $n_x^\sigma(k_y) = \langle \hat{c}^\dagger_{x, k_y;\sigma} \hat{c}_{x, k_y;\sigma} \rangle$ it can be evaluated through the retarded Green's function,
\begin{equation}
n_x^\sigma(k_y) = -\frac{1}{\pi}\int_{-\infty}^{\infty} \mathrm{d}\omega ~n_F(\omega) {\rm Im}\,G_{xx}^{\sigma\sigma}(\omega+i 0^+,k_y),
\label{denSpec}
\end{equation}
where $n_F$ is the Fermi-Dirac distribution. The spectral function at the point $(x,k_y)$ is defined as  $\rho_x^\sigma(\omega,k_y) \equiv -\frac{1}{\pi}{\rm Im}\,G_{xx}^{\sigma\sigma}(\omega+i 0^+,k_y)$. Using the relation 
\begin{equation}n_F(\omega) = \frac{1}{e^{\beta \omega} +1 } = \frac{1}{2} + \frac{1}{\beta} \sum_n \frac{1}{i\omega_n -\omega},
\end{equation}
where $\omega_n = (2n+1)\pi/\beta$ [S1], we have 
\begin{equation}
n_x^\sigma(k_y) =  \frac{1}{2}  \int_{-\infty}^{\infty} d\omega ~\rho_x^\sigma(\omega,k_y) + \frac{1}{\beta} \sum_n \int_{-\infty}^{\infty} d\omega ~ \frac{\rho_x^\sigma(\omega,k_y)}{i\omega_n -\omega} 
 =   \frac{1}{2} +   \frac{1}{\beta} \sum_n G_{xx}^{\sigma\sigma}(i\omega_n,k_y).
\end{equation}
This shows the relation between the number of particles and Matsubara Green's function. The compressibility follows as

\begin{equation}
\kappa_x^\sigma(k_y)
=\frac{\partial}{\partial\mu}\left[\frac{1}{2}+\frac{1}{\beta}\sum_n G^{\sigma\sigma}_{xx}(i\omega_n,k_y)\right]_{\beta,V}
\label{comp2}
\end{equation}
Performing the derivative yields
\begin{equation}
\kappa_x^\sigma(k_y)=-\frac{1}{\beta}\sum_{n,x',\sigma'}G^{\sigma\sigma'}_{x,x'}(i\omega_n,k_y)\left[\frac{\partial}{\partial\mu}G^{-1}(i\omega_n,k_y)\right]^{\sigma'\sigma'}_{x'x'}G^{\sigma'\sigma}_{x'x}(i\omega_n,k_y)
\approx-\frac{1}{\beta}\sum_{n,x',\sigma'}G^{\sigma\sigma'}_{xx'}(i\omega_n,k_y)G^{\sigma'\sigma}_{x'x}(i\omega_n,k_y),
\label{comp3}
\end{equation}
where we assumed that the selfenergy depends only weakly on the chemical potential $\partial \Sigma/\partial \mu \ll 1$. 

The equivalence of the compressibility and the spectral function is shown as follows. 
Using the Dyson equation Eq.~\eqref{denSpec} can be rewritten for $T=0$ as
\begin{equation}
n_x^\sigma(k_y)=-\frac{1}{\pi}\int_{-\infty}^0\mathrm{d}\omega\,\text{Im}\left[\frac{1}{\omega+0^++\mu-H(k_y)-\Sigma(\omega)}\right]_{xx}^{\sigma\sigma}=-\frac{1}{\pi}\int_{-\infty}^\mu\mathrm{d}\omega\text{Im}\left[\frac{1}{\omega+0^+-H(k_y)-\Sigma(\omega-\mu)}\right]_{xx}^{\sigma\sigma}
\end{equation}
where we performed the substitution $\omega+\mu\rightarrow\omega$ in the last  equation. The derivative with respect to $\mu$ yields the compressibility
\begin{equation}
\kappa_x^\sigma(k_y)=-\frac{1}{\pi}\text{Im}\left[\frac{1}{\mu-H(k_y)-\Sigma(0)}\right]_{xx}^{\sigma\sigma}+\mathcal{O}\left(\frac{\partial\Sigma}{\partial\mu}\right).
\end{equation}
The first term corresponds to the spectral function at Fermi level $\rho_x^\sigma(0,k_y)$.

\section{$\mathbb{Z}_2$ index}
The $\mathbb{Z}_2$ index can be  represented  via the extended Matsubara Green's function,
\begin{equation}
P_2 = \frac{\epsilon^{\mu\nu\rho\lambda\sigma}}{120 (2\pi)^3} \int d\omega d{\bm k} \int_{-1}^{1}\int_{-1}^{1} d\bm{\theta}~{\rm{Tr}} \left[ G \partial_\mu G^{-1} G  \partial_\nu G^{-1} G  \partial_\rho G^{-1} G  \partial_\lambda G^{-1} G  \partial_\sigma G^{-1} \right]  = 0 ~{\rm or}~ 1/2~ {\rm (mod~ integer)},
\end{equation}
where $\bm k = (k_x, k_y)$ is the momentum of the physical dimensions and $\bm \theta = (\theta_1, \theta_2)$ is the momentum in the extended dimensions [S2]. The indices $\mu, \nu, \rho, \lambda, \sigma$ run over $\omega, k_x, k_y, \theta_1, \theta_2$. The extended Green's function satisfies $G(\bm k, \bm\theta =0, i\omega) = G(\bm k, i\omega)$ and $G(\bm k, \bm\theta, i\omega) =  \mathcal{T} G(-\bm k, -\bm\theta, i\omega)^T  \mathcal{T}^\dagger$, where $\mathcal{T}  = i \sigma_y \mathcal{C}$. The latter condition is constrained by the time-reversal symmetry. Here, $\mathcal{C}$ is the charge conjugation operator and $\sigma_y$ is the Pauli matrix. Additionally, the Green's  functions at the boundary $\theta_x = \pm 1$ and  $\theta_y = \pm 1$ are fixed to some reference value so that $\theta_x$ and $\theta_y$ constitute a two-dimensional torus (note that the Green's function should be gapped, if $P_2$ is to be quantized).

Now, we introduce the spin-flip transformation for the B sublattice (see Eq.~(6) in the main text). By treating the sublattice index as a pseudospin, the transformation (denoted $U_{\bm i}$, where $\bm i $ is the position of the unit cell) is exactly a global transformation for the internal degrees of freedom (i.e., the spin and the pseudospin), which is position independent, i.e.,  $U_{\bm i} = U $.   The new basis is defined as $\hat{\bm d}_{\bm i} = U \hat{\bm c}_{\bm i}$, and $\hat{\bm d}^\dagger_{\bm i} =  \hat{\bm c}^\dagger_{\bm i} U^\dagger$. Thus we have $\hat{H}_0 = \sum_{\bm i \bm j} \hat{c}_{\bm i}^\dagger h_{\bm i \bm j} \hat{c}_{\bm j} = \sum_{\bm i \bm j} \hat{d}_{\bm i}^\dagger U h_{\bm i \bm j} U^\dagger \hat{d}_{\bm j}$.  In momentum space,  after Fourier transformation, we have  $\hat{\bm d}_{\bm k} = U \hat{\bm c}_{\bm k}$, and $\hat{\bm d}^\dagger_{\bm k} =  \hat{\bm c}^\dagger_{\bm k} U^\dagger$, which shows, that the transformation is  also momentum-independent.  The system is still time-reversal symmetric. But in the new basis, time-reversal has the new representation matrix  $\mathcal{T}^\prime =U \mathcal{T} U^\dagger$.

In the new basis, we have $\tilde{G}(\bm k, i\omega) \equiv \langle T_\tau \hat{\bm d}_{\bm k}(\tau) \hat{\bm d}^\dagger_{\bm k} (0)\rangle = U {G}(\bm k,  i\omega) U^\dagger$.  Similarly, we extend the Green's function to 4+1 dimensions, by setting  $\tilde{G}(\bm k, \bm\theta, i\omega) \equiv U {G}(\bm k, \bm\theta, i\omega) U^\dagger$. Then the Green's function satisfies
\begin{equation}
\tilde{G}(\bm k, \bm\theta, i\omega)  = U \mathcal{T} G(-\bm k, -\bm\theta, i\omega)^T  \mathcal{T}^\dagger U^\dagger = \mathcal{T}^\prime \tilde{G}(-\bm k, -\bm\theta, i\omega)^T   \mathcal{T^\prime}^\dagger.
\end{equation}
Since the value of the trace does not change after a unitary transformation of the matrix, we obtain 
\begin{equation}
P_2 = \frac{\epsilon^{\mu\nu\rho\lambda\sigma}}{120 (2\pi)^3} \int d\omega d{\bm k} \int_{-1}^{1}\int_{-1}^{1} d\bm{\theta}~{\rm{Tr}} \left[ \tilde G \partial_\mu \tilde G^{-1} \tilde G  \partial_\nu \tilde G^{-1} \tilde G  \partial_\rho \tilde G^{-1} \tilde G  \partial_\lambda \tilde G^{-1} \tilde G  \partial_\sigma \tilde G^{-1} \right].
\end{equation}
The right part is exactly the $\mathbb{Z}_2$ index for the system in the new basis with the time reversal symmetry $\mathcal{T}^\prime$. The system in the new basis, decouples into two subsystems, i.e. their respective particle number is conserved, in which case, the $\mathbb{Z}_2$ index can be expressed as the difference of the two Chern numbers of the two spin subsystems.

\section{Local spin Chern marker}
\begin{figure}[ht]
\includegraphics[width=.7\columnwidth]{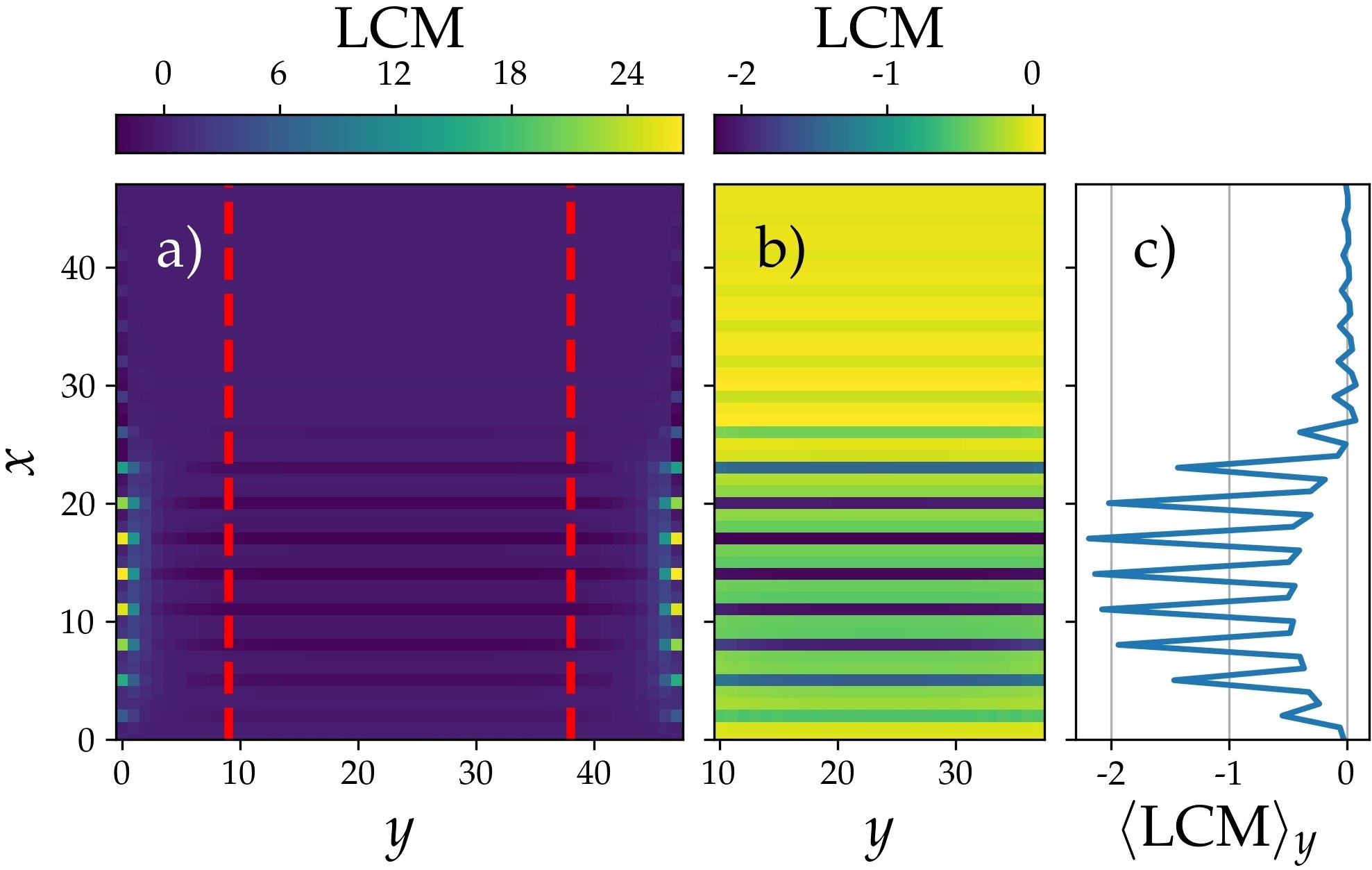}
\caption{Local Chern marker of a) the full lattice, b) the cut-off lattice, as depicted in a) by the dashed red lines, and c) the average value of b) along the $y$-direction for  $U=0$, $\lambda_L=0$, and $\lambda_R=2.5$ with OBC in $x$-direction and PBC in $y$-direction. Edge effects in the $y$-direction emerge through the ill-defined position operator $\hat{y}$.}
\label{mark}
\end{figure}
Equation (12) in the main text corresponds to the operator representation of the original formula [S3] of the local Chern marker. Note that for PBC the position operators are ill-defined and one is left with boundary effects. These boundary effects must be cut-off when we consider the bulk value of the LCM. In Fig.~\ref{mark} we show the LCM for $U=0$, $\lambda_L=0$, and $\lambda_R=2.5$ for a) the full lattice, b) the cut-off lattice, corresponding to the region in a) enclosed by the dashed red lines, and c) the average of b) along the $y$-direction. 

\begin{center}
--------------------------
\end{center}

\small{
[S1] G. D. Mahan, \textit{Many-Particle Physics}  (Springer, Boston, MA, 2000).

[S2] Z. Wang, X.-L. Qi, S.-C. Zhang, Phys. Rev. Lett. {\textbf 105}, 256803 (2010).

[S3] R. Bianco and R. Resta, Phys. Rev. B \textbf{84}, 241106 (2011).
}